\documentclass[aps,prl,epsf,twocolumn,superscriptaddress,longbibliography]{revtex4-1}
\usepackage[pdftex]{graphicx}
\usepackage{dcolumn}
\usepackage{bm}

\begin{document}


\def\PTO{PbTiO$_3$}
\def\STO{SrTiO$_3$}

\def\comment#1{{\large\textsl{#1}}}
\def\degree {{$^\circ$}}
\def\degrees{{$^\circ$}}
\def\eq#1{{Eq.~(\ref{eq:#1})}}
\def\fig#1{{Fig.~\ref{fig:#1}}}
\def\sec#1{{Sec.~\ref{sec:#1}}}
\def\inv{^{-1}}
\def\micron {\hbox{$\mu$m}}
\def\microns{\micron}
\def\Ref#1{{Ref.~\onlinecite{#1}}}  
\def\tab#1{{Table~\ref{tab:#1}}}
\def\tauOne{\tau^{(1)}}
\def\tVec{\hbox{\bf t}}
\def\thetaDet{\theta_{DET}}

\def\qvec{{\vec q}}
\def\pvec{{\vec p}}
\def\Avec{{\vec A}}
\def\qhat{{\hat q}}
\def\qperphat{{\hat q_\perp}}
\def\ekpq{{E_{\kvec+\qvec}}}
\def\ek{{E_{\kvec}}}
\def\Omegabar{{\bar\Omega}}
\def\omegabar{{\bar\omega}}
\def\omegap{{\omega_p}}
\def\kf{{k_F}}
\def\kappaf{{\kappa_F}}
\def\mone{{-1}}
\def\re{{\rm{Re\,}}}
\def\im{{\rm{Im\,}}}
\def\twopi{{2 \pi}}
\def\wpm{w_\pm}
\def\FWlindhard{Appendix~A}
\def\lindhardTrans{Appendix~B}
\def\ftr{{f^{tr}}}
\def\PN{Pines and Nozi{\`e}res}
\def\Bohm{B{\"o}hm}
\def\Nifosi{Nifos{\'\i}}
\def\prin{{\cal P}}

\def\half{{1/2}}
\def\minusHalf{{-1/2}}
\def\threeHalves{{3/2}}
\def\minusThreeHalves{{-3/2}}

\title{Local atomic geometry and Ti 1s near-edge spectra in PbTiO$_3$ and SrTiO$_3$}

\author{Eric Cockayne}
\email{eric.cockayne@nist.gov}
\affiliation{
Materials Measurement Science Division, 
Material Measurement Laboratory, 
National Institute of Standards and Technology, 
Gaithersburg, Maryland 20899 USA}

\author{Eric L. Shirley}
\email{eric.shirley@nist.gov}
\affiliation{
Sensor Science Division,
Physical Measurement Laboratory,
National Institute of Standards and Technology,
Gaithersburg, Maryland 20899 USA}

\author{Bruce D. Ravel}
\affiliation{
Materials Measurement Science Division,
Material Measurement Laboratory,
National Institute of Standards and Technology,
Gaithersburg, Maryland 20899 USA}

\author{Joseph C. Woicik}
\affiliation{
Materials Measurement Science Division,
Material Measurement Laboratory,
National Institute of Standards and Technology,
Gaithersburg, Maryland 20899 USA}

\date{\today}

\begin{abstract}

 We study Ti 1s near-edge spectroscopy in PbTiO$_3$ at various 
temperatures above and below its tetragonal-to-cubic phase transition, 
and in SrTiO$_3$ at room temperature.
{\em{Ab initio}} molecular dynamics (AIMD) runs on 80-atom supercells are used
to determine the average internal coordinates and their fluctuations. 
We determine that one vector local order parameter is the dominant contributor to 
changes in spectral features: the displacement of the Ti ion with respect to its
axial O neighbors in each Cartesian direction, as these displacements enhance the 
cross section for transitions to E$_{\rm g}$-derived core-hole exciton levels.
Using periodic five-atom structures whose relative Ti-O displacements match 
the root-mean-square values from the AIMD simulations, and core-hole Bethe-Salpeter equation 
(BSE) calculations, we quantitatively predict the respective Ti 1s near-edge spectra.  
Properly accounting for atomic fluctuations greatly improves the agreement
between theoretical and experimental spectra.
The evolution of relative strengths of spectral features vs temperature 
and electric field polarization vector are captured 
in considerable detail.  
This work shows that local structure can be characterized from first-principles 
sufficiently well to aid both the prediction and the interpretation of near-edge spectra.  


\end{abstract}


\maketitle
\thispagestyle{empty}

\section{Introduction}
\label{sec:intro}

Perovskite ABO$_3$ compounds are a family of materials of technological interest, 
in part because of their usefulness as dielectric and piezoelectric materials.
A variety of phase transitions are seen in perovskite oxides, from
ferroelectric transitions in BaTiO$_3$ and PbTiO$_3$ to antiferrodistortive
transitions in SrTiO$_3$ and more complicated transitions in CaTiO$_3$ and
PbZrO$_3$.  
Additionally, {\em relaxor} behavior\cite{Cross1994,Bokov2006,Cowley2011} 
(a broad, frequency-dependent dielectric maximum vs temperature,
not associated with a paraelectric-ferroelectric phase transition)
can occur in perovskite solid solutions
such as Pb[Mg$_{1/3}$Nb$_{2/3}$]O$_3$ (PMN).

Many of the practical properties of these materials are due to
the displacement of ions in response to 
temperature, electric fields, stress, etc.  Response functions can 
also
be related to fluctuations and correlations of the ionic positions over time and/or
space~\cite{Rabe1998,Ponomareva2008}.  To clarify structure-property relationships, it is therefore 
important to fully characterize the ionic positions as a function of 
temperature, etc.,  not only the average positions, 
but also their fluctuations around the average over time and/or unit cells.
Such distributions can reveal interesting features that can dramatically affect
a material's properties, such as characteristic deviations of local structure from the average.
While a difference between local atomic geometries and average
properties is naturally expected in complex perovskites where different species 
occupy the same crystallographic site, pioneering work of 
Comes {\it et al.}\cite{Comes1968} gave evidence that even simple perovskites 
such as BaTiO$_3$ and KNbO$_3$ in their cubic phases can exhibit correlated 
off-centering of their B ions ({\it e.g.}, Ti or Nb).


There are many theoretical methods to determine local atomic geometries in materials.  
Density-functional theory (DFT) ground-state calculations are direct, but 
are limited regarding the sizes of systems that can be treated.
Also, DFT calculations are usually restricted to the zero-temperature case, 
and are subject to systematic 
error because of approximate treatments of exchange and correlation.  
To overcome the DFT zero-temperature problem, DFT-based {\em {ab initio}} molecular dynamics (AIMD) calculations
allow finite-temperature fluctuations of ionic coordinates to be incorporated 
in the Born-Oppenheimer limit.   
However, the sizes of treatable systems may be even further restricted,  
and one must cover a sufficiently long simulation time to equilibrate
to representative conditions at a given temperature.
A molecular dynamics calculation that uses first-principles-based effective Hamiltonians or
force-field models allows for effects involving greater length and time scales to be investigated,
but its predictions are generally not as accurate as those obtained using AIMD.

There are also many experimental methods to determine
local atomic geometry~\cite{Billinge2007}.  
For example, x-ray absorption fine structure spectroscopies 
[near-edge x-ray absorption fine structure (NEXAFS), 
also known as x-ray absorption near-edge structure (XANES), 
and extended x-ray absorption fine structure (EXAFS)]
can furnish a wealth of information via interference effects 
because of a core-level photoelectron's multiple scattering from 
atomic sites~\cite{Rehr05}.  Such core level spectroscopies 
are element-specific, and the excitonic ``pre-edge" 
features (involving 1s-to-3d transitions from the ground state 
to a bound core-hole exciton state) are strongly affected 
by variations in the B-site (e.g., Ti) position with respect to its nearest neighbors.
As an example, this effect was previously demonstrated in strained, 
thin-film ferroelectric SrTiO$_3$~\cite{Woicik2007}.  
In that work, x-ray diffraction and DFT determination of the unit cell size and 
internal atomic coordinates agreed 
with respect to
the detailed crystal structure, 
and theoretical and experimental NEXAFS calculations indicated similar 
degrees of Ti off-centering in the TiO$_6$ cages.  
Bulk calculations confirmed selection rules for excitation into
$T_{\rm 2g}$ and E$_{\rm g}$ core-hole exciton levels, and these rules were 
violated in a fashion that is consistent with 
average displacements and 
thermal fluctuations 
of the Ti coordinate relative to its O$_6$ cage.  

Lead titanate (PbTiO$_3$) is a compound of interest, in part 
because it is an endmember of important solid solutions,
such as lead-zirconium titanates (PZT).
It has a single tetragonal/cubic transition at $T=763~{\mathrm{K}}$, 
with the cubic phase existing at higher temperatures.  
At room temperature, the Ti 1s near-edge spectrum 
in PbTiO$_3$ features an E$_{\rm g}$ peak 
near 4970.5~eV that is strongly enhanced 
compared to the T$_{\rm 2g}$ peak 
for x-ray photon electric field polarization 
along the tetragonal axis~\cite{Woicik2007}.  
This is a hallmark signature of Ti off-centering relative
to the surrounding O$_6$ octahedra in the tetragonal phase.
However, the peak is still enhanced in the cubic phase
above the transition temperature, whereas Bethe-Salpeter equation 
(BSE) calculations that assume 
only {\em{average}} cubic perovskite structure fail to predict this enhancement.  
Instead, thermal fluctuations must be taken into account~\cite{Vedrinskii1998}.  
Related EXAFS studies~\cite{Sicron1994,Yoshiasa2016} also indicate off-centering of 
Pb and Ti in PbTiO$_3$ above the transition temperature.

In this work, we combine AIMD and a BSE
treatment of the Ti 1s near-edge  spectrum 
in a fashion that addresses the thermal fluctuations.  
We find that the important fluctuations for a given temperature can 
be effectively reproduced by a single periodic five-atom cell.
BSE calculations are performed using these cells and 
compared with measured spectra~\cite{Vedrinskii1998}
to investigate the local structure in {\PTO} as a function of
temperature and in {\STO} at room temperature.

In what follows, we review the background and theoretical methodology (i.e., AIMD and BSE calculations).  
Aspects of measurements of experimental spectra to which we compare results 
are given in more detail elsewhere\cite{Vedrinskii1998}.
We then present our results, comparing calculated and measured spectra, 
and present our conclusions, primarily that
our present accounting of ionic fluctuations
greatly improves agreement between theoretical
and measured spectra in the pre-edge region.

\section{Background and Methodology}
\label{sec:methods}

\subsection{Underlying Physics}

Let a Ti 1s electron's coordinate relative to the nucleus be denoted by ${\bf r}$.   
Then the effective light-matter interaction operator responsible for a core excitation 
can be approximated as
\begin{equation}
\hat{O} \approx ({\bf e}\cdot{\bf r})+\frac{i}{2}({\bf e}\cdot{\bf r})({\bf q}\cdot{\bf r}).
\end{equation}
Here, ${\bf e}$ is the photon's electric-field polarization vector, and ${\bf q}$ is its wave vector.  
Thus, the vectors ${\bf q}$ and ${\bf e}$ determine how the two terms respectively allow 
electric dipole (E1) and quadrupole (E2) transitions at the Ti 1s near-edge.  
As a consequence, a variety of experimental parameters can be varied to 
``tease out" and assign various spectral features, 
depending on the orientations of ${\bf e}$ and ${\bf q}$ 
relative to the crystallographic directions, 
as well as temperature, film thickness, polarization, strain and flexure.  
Near-edge spectra in SrTiO$_3$ and PbTiO$_3$ reveal electric-dipole
and electric-quadrupole transitions to Ti 4p, 3d and mixed 4p-3d states.
Weak, crystal-field-split ``pre-edge" features are 
attributed to 1s$\rightarrow$3d transitions 
and are small compared 
to the main ``edge jump" at the onset of the Ti 4p continuum states 
around and above 4970 eV.

For octahedral Ti site symmetry, there are two leading pre-edge features, 
around 4968.5~eV and 4970.5~eV, because the T$_{\rm 2g}$ and E$_{\rm g}$ orbitals are
separated by the ligand-field splitting.  
These are understood as excitation to 3d level on the same site as the 1s hole, 
while the next features around 4974~eV could arise from excitation to 3d levels on 
nearby Ti sites because of hybridization with the on-site 4p states mediated by the 
O 2p states~\cite{Vedrinskii1998,deGroot2009}.  
In the presence of a Ti 1s core hole, 
an electron in any of the on-site orbitals would be bound to the Ti site,
so that the electron-core hole pair constitutes a core-hole exciton.  
Pseudo Jahn-Teller splittings and site symmetry lower than cubic symmetry 
can split the otherwise threefold and twofold degeneracies further.  
The edge jump is chiefly due to the E1 term, 
which can greatly enhance pre-edge features when Ti-site inversion symmetry is broken.  
In that case, low lying Ti 4p-derived and 3d-derived E$_{\rm g}$ states, 
which are close in energy, mix, so that the latter become 
more strongly accessible via the E1 term.   
Otherwise, the E2 term alone can give rise to the weak pre-edge features.  
Selection rules imply that 3d lobes should have projections 
in the plane of ${\bf e}$ and ${\bf q}$ but not along either of these vectors.  
As an example, the antiferrodistortive tetragonal phase of 
\STO~should affect the E2 selection rules regarding observation of 
$d_{xy}$-derived vs $d_{x^2-y^2}$-derived exciton levels
when ${\bf e}$ and ${\bf q}$ are both in the basal plane 
because of TiO$_6$ cage rotations.  
However, fluctuations of the cage rotational angles about their 
average values can be substantial, 
further weakening the selection rules.
As examples of past work in the near-edge spectroscopy of 
simple perovskites, Ti off-centering in {\PTO} was
analyzed by Vedrinskii~{\em{et al.}}~\cite{Vedrinskii1998}, 
and Ravel~{\em{et al.}}~\cite{Rav}, and  Ti off-centering in {\STO}
by several groups~\cite{Nozawa2005,Woicik2007}.  
Other effects of symmetry were studied 
theoretically and experimentally 
in non-perovskite compounds containing TiO$_6$ cages, such as
rutile and anatase~\cite{Kotani,Grunes1983,Durmeyer1990}.  

\subsection{Ab initio molecular dynamics (AIMD) calculations}

\begin{figure}[h]
\includegraphics[width=85mm]{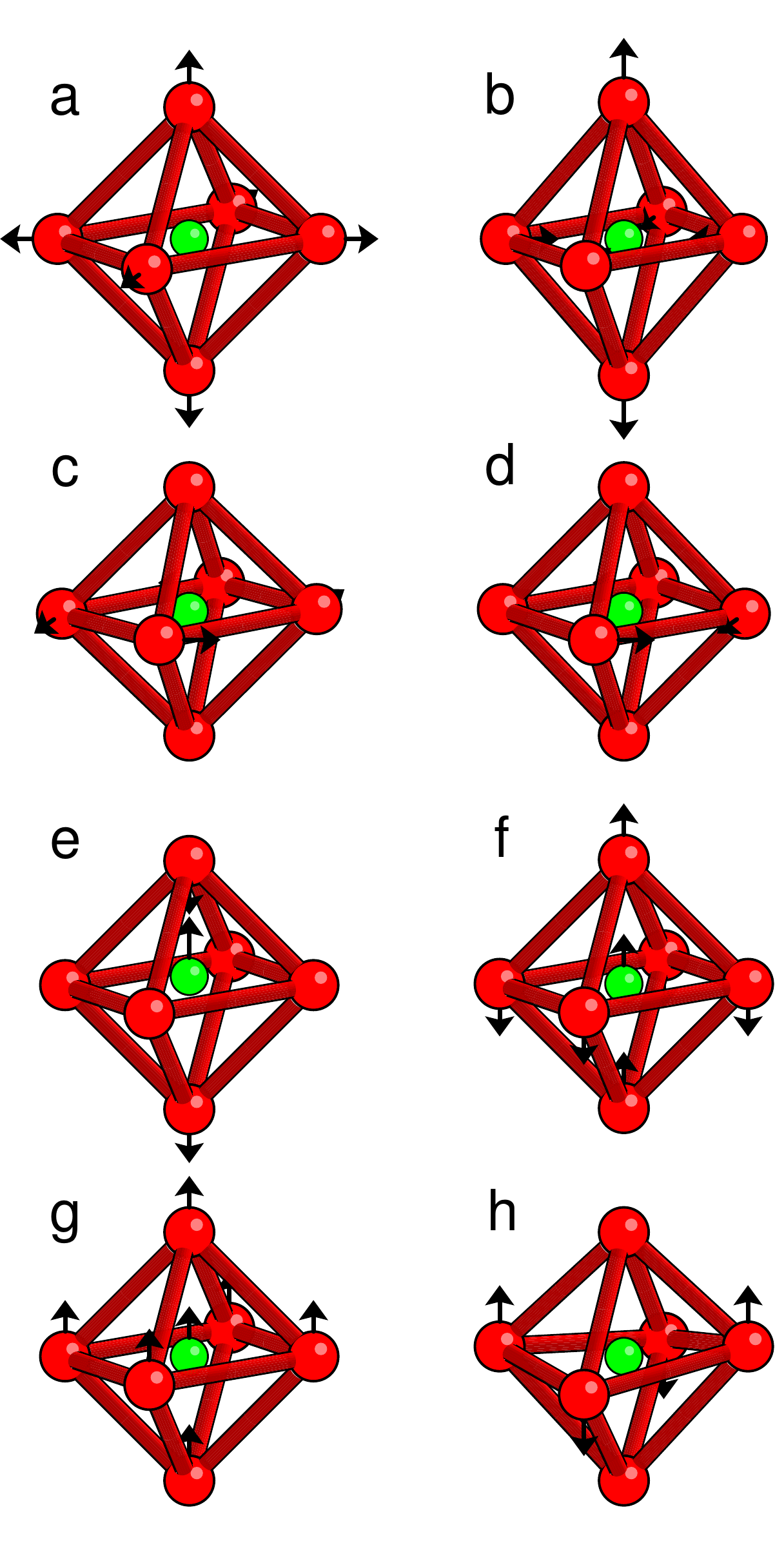}
\caption{One component of each irreducible representation for local TiO$_6$
distortions in a cubic titanate perovskite: (a) a$_{\rm 1g}$;
(b) e$_g(z)$; (c) t$_{\rm 1g}(z)$; (d) t$_{\rm 2g}(z)$;
(e) t$_{\rm 1u}^{(1)}(z)$; (f) t$_{\rm 1u}^{(2)}(z)$;
(g) t$_{\rm 1u}^{(3)}(z)$;
(h) t$_{\rm 2u}$(z). Some ``away-pointing" arrows are obscured.}
\label{fig:irreps}
\end{figure}

From the theory in \sec{methods}(A), the physics that dominates
the Ti near-edge in PbTiO$_3$ and SrTiO$_3$ involves Ti ions
and their nearest O neighbors.
The pertinent
local environment thus features 
seven atoms: the Ti itself and an octahedron of six surrounding O ions.  
Under $O_h$ symmetry, 
the 21 corresponding vibrational degrees of freedom transform into
irreproducible representations (irreps) as follows: a$_{\rm 1g}$ + 
e$_{\rm g}$ +
t$_{\rm 1g}$ + t$_{\rm 2g}$ + 3 t$_{\rm 1u}$ + 
t$_{\rm 2u}$.
One displacement pattern for each irrep is given in~\fig{irreps}.
and the corresponding normalized full displacement patterns are given in 
Supporting Information Table S1.
In the most important local displacement pattern in this work
t$_{\rm 1u}^{(1)}$(z), the central Ti ion moves 
proportionally to 
$+(\sqrt{2/3})~c$ along $z$ and
the the two neighboring O ions in
the $\pm z$ direction move 
proportionally to 
$-(\sqrt{1/6})~c$ along $z$, 
where $c$ is
the lattice parameter of the primitive five-atom perovskite cell 
along $z$ (see~\tab{modes}).
For the corresponding t$_{\rm 1u}^{(1)}(x)$ and t$_{\rm 1u}^{(1)}(y)$ patterns,
$z$ is replaced by $x~(y)$ and $c$ is replaced by $a$.
For simplicity, the same mode labels are used here for cubic and for tetragonal PbTiO$_3$;
tetragonal cell modes are identified with the cubic irrep from which they are derived.

\begin{table*}[t]
\caption{Atomic displacements corresponding to unit amplitude of the
t$_{\rm 1u}^{(1)}(z)$ degree of freedom for the TiO$_6$ cluster in a perovskite titanate.
Entries that are zero are left blank.
The unit-amplitude t$_{\rm 1u}^{(1)}(x)$ and t$_{\rm 1u}^{(1)}(y)$ displacement patterns
can be found be cyclic permutation and by replacing the $z$ direction 5-atom primitive perovskite cell lattice
paramter $c$ with the corresponding lattice parameter in the $x$ or $y$ direction.
Atomic displacement patterns for the other TiO$_6$ cluster degrees of freedom can be found in the
Supporting Information Table S1.}

\vspace*{0.3in}

\begin{tabular}{ccc|ccc|ccc|ccc|ccc|ccc|ccc}
\hline
          & Ti &    &    & O$_{+x}$ & &   & O$_{-x}$ &  &    & O$_{+y}$ & & & O$_{-y}$ &    &  & O$_{+z}$ &   &  & O$_{-z}$ &  \\
              $x$ &  $y$ &  $z$ &  $x$ &  $y$ &  $z$ &  $x$ &  $y$ &  $z$ &  $x$ &  $y$ &  $z$ &  $x$ &
 $y$ &  $z$ &  $x$ &  $y$ &  $z$ &  $x$ &  $y$ &  $z$ \\
  &  & $(2/\sqrt{6}) c$  &  &  &    &  &  &    &  &  &    &  &  &    &  &  & $(-1/\sqrt{6}) c$ &  &  & $(-1/\sqrt{6}) c$ \\
\hline
\label{tab:modes}
\end{tabular} \end{table*}

Ab initio molecular dynamics~\cite{Payne1992} using DFT at the 
generalized-gradient-approximation (GGA) level with Hubbard $U$ corrections
was used to compute the time varying atomic coordinates, ${x_{\mu i}(t)}$.  
Here, $t$ denotes time, $\mu$ labels one of the above 21 local modes, and
$i$ the position (unit cell).
Simulations were run 
for {\PTO} at 300~K, 600~K, and 900~K.  
For all runs, we used 80-atom supercells with lattice vectors (0,2,2), (2,0,2), and (2,2,0) in
terms of the lattice vectors of the corresponding 5-atom perovskite perovskite
cell.
The MD was run using the Vienna Ab initio Simulation Package (VASP) code~\cite{Kresse96,disclaimer}.    
The simulation time step was 1.25~fs, 
and the total simulation time was 2.5~ps (2000 time steps).  
Langevin dynamics were used, with a frequency of 2~THz.  The final runs were performed after
a series of initialization runs totaling 2~ps.  Such a simulation time is relatively short
compared to other AIMD studies of PbTiO$_3$~\cite{Srinivasan2003,Fang2015}, because we 
enforced high convergence with respect to k-points and plane-wave cutoff energy.  
Nevertheless, the 2.5~ps run is sufficient for convergence of atomic coordinate averages and 
fluctuations, as found by consistency of the results of the final run with those of the previous
1.25~ps initialization run.

It is well known that the GGA 
gives a $c/a$ ratio for PbTiO$_3$ that is much larger than experiment~\cite{Metri2005}.  
We have found empirically that the PBEsol~\cite{Perdew05} GGA
with a Hubbard correction of $U=2.3~{\mathrm{eV}}$ for oxygen p-states and
an artificial pressure of $-$21~kbar ($-$2.1~GPa) reproduces 
low-temperature structures of other perovskite titanates such as BaTiO$_3$, CaTiO$_3$, and
SrTiO$_3$ quite well, but the $c/a$ ratio of PbTiO$_3$ remains too high.
(It has recently been suggested that the inclusion of van der Waals forces, neglected here, 
can strongly affect the calculated $c/a$ ratio~\cite{Berland2014}.)
On the other hand, we find that a room-temperature AIMD run for PbTiO$_3$ with the
above parameterization, and with the unit cell fixed at the experimental one,
yields an average structure with ionic positions within {0.05~\AA} of the experimental values.
Therefore, each AIMD run was performed with the primitive lattice parameters fixed at the 
experimental values at the corresponding temperature~\cite{Mabud1979,Wyckoff1964}.  
The lattice constants used are presented in Table~\ref{tab:latparam}.

\begin{table}
\caption{Experimental lattice parameters for SrTiO$_3$ and PbTi$O_3$\cite{Mabud1979,Wyckoff1964}
used in this work (in~\AA).}

\begin{tabular}{ccccccc}
\\
\hline
     & \phantom{askdf} & SrTiO$_3$ & \phantom{askdf}  & &   PbTiO$_3$ & \\
$T({\mathrm{K}})$  &&  293     && 300  &  600      & 900 \\
$a$  &&  3.9051   && 3.901 & 3.927 & 3.980 \\
$c$  &&  3.9051   && 4.157 & 4.099 & 3.980  \\
\hline
\label{tab:latparam}
\end{tabular}
\end{table}

Defining $\langle Q\rangle_{t}$ as an average of quantity $Q$ over time and  
$\langle Q\rangle_{i}$ as an average over sites,
we define the mode average of a local coordinate by its mean value, namely 
\begin{equation}
\overline{x_{\mu}} = \langle\langle x_{\mu i}(t)\rangle_{i}\rangle_{t},
\end{equation}
and the root-mean-square local coordinate as 
\begin{equation}
\sigma({\mu}) = (\langle\langle x_{\mu i}^2(t)\rangle_{i}\rangle_{t})^{1/2}.
\end{equation}
We also define the time-and-space deviation of a coordinate
from its mean value by $x^{\prime}_{\mu i}(t) = x_{\mu i}(t) - \overline{x_{\mu}}$.
Most of the mode averages $\overline{x_{\mu}}$  are zero by
symmetry.  However, in the tetragonal PbTiO$_3$ phase, the 
average amplitudes of the polar t$_{\rm 1u}(z)$ modes are nonzero.

\subsection{Bethe-Salpeter equation (BSE) calculations}


 In \sec{methods}(B), we describe the 21 degrees of freedom of a TiO$_6$ unit,
and how their averages and root-mean-squares were calculated via AIMD.
Before discussing the details of our Bethe-Salpeter equation calculations,
we determine which of these 21 degrees of freedom are important.
Rigid translation of the TiO$_6$ cage (t$_{\rm 1u}^{(3)}$ modes)
has no effect on the absorption spectrum, while rigid rotation (t$_{\rm 1g}$) serves 
only to change the polarization selection rules for E2 transitions. 
Of the remaining 15 degrees of freedom, isotropic breathing (a$_{\rm 1g}$) has little impact on the spectrum,
while the two e$_g$ breathing modes are responsible for pseudo-Jahn-Teller splitting of the 
1s-to-$E_g$ peak \cite{Tinte2008,Gilmore2010}. 
The t$_{\rm 2g}$, t$_{\rm 1u}^{(2)}$ and 
t$_{\rm 2u}$ modes do not change Ti-O distances to first order. 
Thus, the set of three t$_{\rm 1u}^{(1)}$ modes in which the Ti ion moves relative to its axial O neighbors
are expected to dominate the near edge spectra.
Furthermore, the spectra should be proportional to the
square of the amplitude of the mode.  Thus, averaging over unit cells, the
expected spectra are equivalent to that for a single periodic five-atom cell where the
amplitude of each component of the t$_{\rm 1u}^{(1)}$ displacement pattern is equal to
the corresponding root-mean square value over all cells.  
Explicit expressions for the effective atomic coordinates, valid for 
ATiO$_3$ perovskites 
of cubic and tetragonal symmetry, are given in Table~\ref{tab:nexafscell}.
The A$_z$ and O$_z$ displacements in Table~\ref{tab:nexafscell} under tetragonal
symmetry are insignificant (for near-edge spectra purposes), but are set at their values
determined from the average AIMD structure.

\begin{table*}[t]
\caption{Coordinates of 5-atom structures for use in the
Bethe-Salpeter equations to compute near-edge spectra of cubic or tetragonal
simple ATiO$_3$ perovskites, given root-mean-square fluctuations associated with
the t$_{\rm 1u}$ local coordinate (\tab{modes}).
The origin is chosen such that the crystallographic
average position of one O atom is (0.5, 0.5, 0.0).  A$_z$ and
O$_z$ coordinates are set at their average values.}
\begin{tabular}{lccc}
\\
\hline \hline
\\
& & $T$=300~K: \\
{\bf{Ion}} & {\bf{\em x}} & {\bf{\em y}} & {\bf{\em z}} \\ \hline
A & 0 & 0 & A$_z$   \\
Ti & 0.5 + ($2/\sqrt{6}$) $\sigma(t_{1u}^{(1)}(x))$ \,\,\,\,  & 0.5 + ($2/\sqrt{6}$) $\sigma(t_{1u}^{(1)}(y))$ \,\,\,\,  & 0.5 + ($2/\sqrt{6}$) $\sigma(t_{1u}^{(1)}(z))$ \\
O  & 0.0 - ($1/\sqrt{6}$) $\sigma(t_{1u}^{(1)}(x))$ & 0.5 &  O$_z$ \\
O  & 0.5 & 0.0 - ($1/\sqrt{6}$) $\sigma(t_{1u}^{(1)}(y))$  & O$_z$ \\
O  & 0.5 & 0.5 & 0.0 - ($1/\sqrt{6}$) $\sigma(t_{1u}^{(1)}(z))$  \\ \hline
\\
\hline \hline
\label{tab:nexafscell}
\end{tabular} \end{table*}

 

In our calculations, we chose to have the c-axis along the $z$ direction.  
Two x-ray photon geometries were considered, 
corresponding to what was measured.  
One geometry had  
${\bf{q}} \parallel (-1,-1,0)$ and $ {\bf{e}} \parallel (0,0,1)$, 
and one geometry had 
${\bf{q}} \parallel (0,-1,1)$ and $ {\bf{e}} \parallel (1,0,0)$.  
Randomization of signs of displacements along each Cartesian direction 
was incorporated 
by averaging spectra with suitable variation of the signs of Cartesian 
components of ${\bf {q}}$ and ${\bf {e}}$.  

The x-ray absorption coefficient can be given as 
\begin{equation}
\mu(E) = - {\mathrm{Im}} \langle \Phi_0 \mid 
\hat{O}^{\dagger} \frac{1}{E+i \Gamma(E) - H} \hat{O} 
\mid \Phi_0 \rangle,
\end{equation}
where $\Phi_0$ denotes the ground state, 
$\Gamma(E)$ accounts for lifetime damping of excited states, 
and $H$ is the effective Hamiltonian for the electron core-hole pair's 
equation of motion, which is the Bethe-Salpeter equation.  
This Hamiltonian includes effects of the band structure, 
electron self-energy damping calculated as described elsewhere~\cite{FisterLiXYZ}, 
core hole lifetime damping~\cite{AMRD137}, 
and the very important screened electron-core hole interaction.  
The Bethe-Salpeter equation solver is described 
by Vinson~{\em{et al.}}~\cite{Vinson2011},  
with various details regarding the calculation of matrix elements
given elsewhere~\cite{KotaniFest}.    

A precursor DFT calculation underlies the BSE calculation.   
This provides Bloch state wave functions and energies.
We used a 200~Ry plane-wave cutoff.  This was necessary 
to describe Ti 3s and 3p states well, 
while simultaneously having Ti 4s and 4p energies 
sufficiently accurate by virtue of a hard pseudopotential for Ti.  
Thus, we used He-like, Ne-like and Hg$^{2+}$-like cores 
for Vanderbilt-type pseudopotentials~\cite{Vanderbilt85}
 for O, Ti and Pb, respectively.   
We used the local-density approximation~\cite{Kohn1965} 
with the Ceperley-Alder functional~l\cite{CA}
as parameterized by Perdew and Zunger~\cite{PZ}.  
The self-consistent charge density was found using  
$4\times4\times4$ Monkhorst-Pack grids~\cite{MP}. 

The BSE calculation include 155 bands total and $10\times10\times10$ grids.  
This was accelerated by use of optimized basis functions described by 
Prendergast and Louie~\cite{Prendergast2009}, 
who improved upon earlier work~\cite{Shirley1996OBF}.  
The electron-core hole 
interaction was screened as described elsewhere~\cite{EDGE}.  
For the electronic dielectric constant, we used 
$\epsilon_{\infty}=7.25$  at  $T=300~{\mathrm{K}}$, 
$\epsilon_{\infty}=7.60$  at  $T=600~{\mathrm{K}}$, 
$\epsilon_{\infty}=8.68$  at  $T=900~{\mathrm{K}}$, 
which averages the values of the three diagonal elements 
of the measured dielectric tensor at each temperature~\cite{Kleemann1986}.    

\section{Results and Discussion}



\begin{table}
\caption{Temperature-dependent average structure of PbTiO$_3$ from AIMD}
\begin{tabular}{lccc}
\\
\hline \hline
\\
& & $T$=300~K: \\
{\bf{Ion}} & {\bf{\em x}} & {\bf{\em y}} & {\bf{\em z}} \\ \hline
Pb & 0.0000 & 0.0000 & 0.1056 \\
Ti & 0.5000 & 0.5000 & 0.5709 \\
O  & 0.0000 & 0.5000 & 0.4982 \\
O  & 0.5000 & 0.0000 & 0.4982 \\
O  & 0.5000 & 0.5000 & 0.0000 \\ \hline
\\
& & $T$=600~K: \\
{\bf{Ion}} & {\bf{\em x}} & {\bf{\em y}} & {\bf{\em z}} \\ \hline
Pb & 0.0000 & 0.0000 & 0.0739 \\
Ti & 0.5000 & 0.5000 & 0.5550 \\
O  & 0.0000 & 0.5000 & 0.4993 \\
O  & 0.5000 & 0.0000 & 0.4993 \\
O  & 0.5000 & 0.5000 & 0.0000 \\ \hline
\\
& & $T$=900~K: \\
{\bf{Ion}} & {\bf{\em x}} & {\bf{\em y}} & {\bf{\em z}} \\ \hline
Pb & 0.0000 & 0.0000 & 0.0000 \\
Ti & 0.5000 & 0.5000 & 0.5000 \\
O  & 0.0000 & 0.5000 & 0.5000 \\
O  & 0.5000 & 0.0000 & 0.5000 \\
O  & 0.5000 & 0.5000 & 0.0000 \\ \hline
\\
\hline \hline
\label{tab:pbtio3a}
\end{tabular} \end{table}

\begin{figure}[h]
\includegraphics[width=85mm]{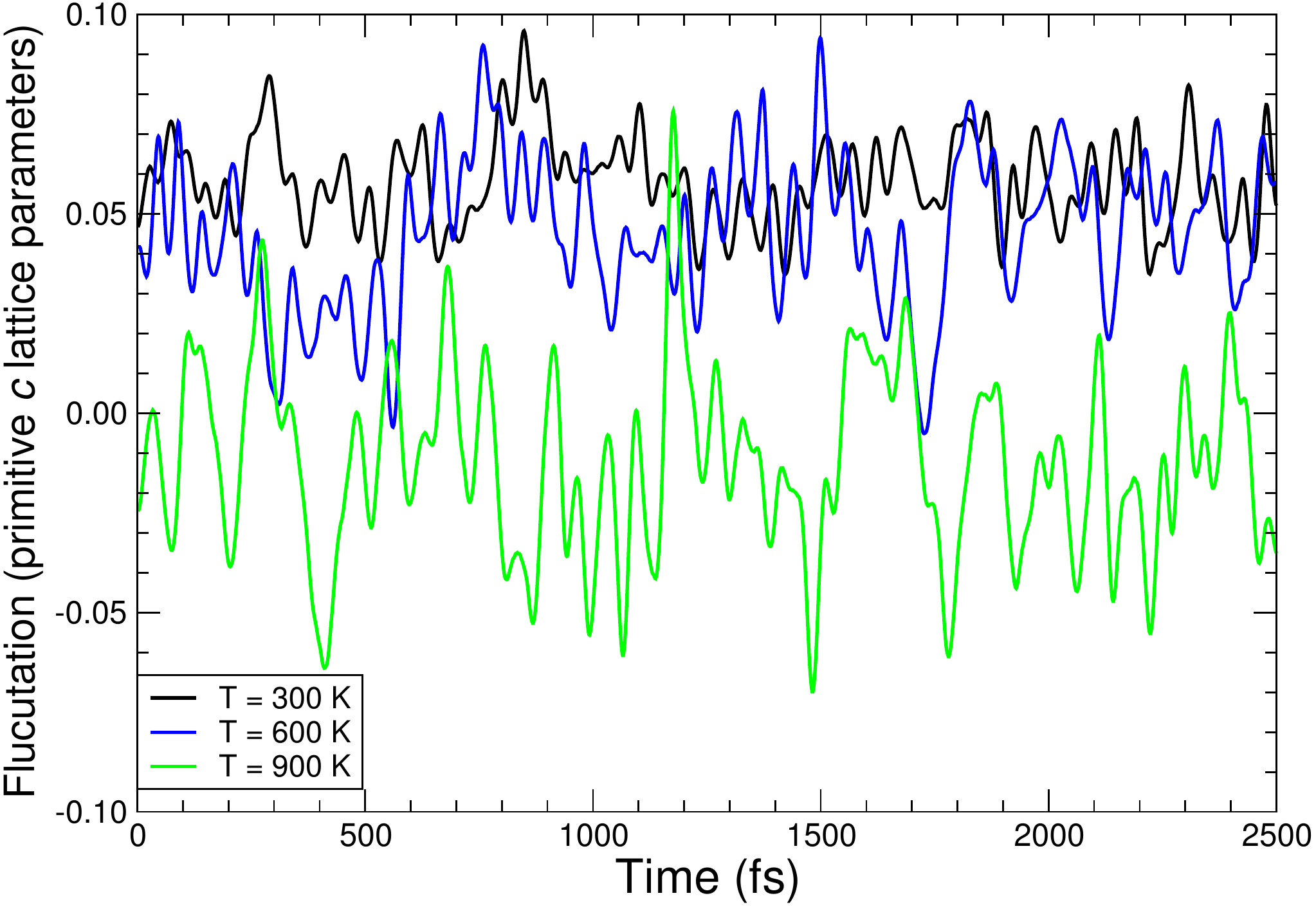}
\caption{t$_{\rm 1u}^{(1)}$(z) fluctuation of one Ti against its $\pm z$
oxygen neighbors vs MD time, at different temperatures.}
\label{fig:tfluck}
\end{figure}

\begin{table}
\caption{Root-mean-square fluctuations
$\sigma({\mu})$ of the t$_{\rm 1u}^{(1)}$ modes of
PbTiO$_3$ as a function of temperature, as computed by {\em{ab initio}} molecular dynamics.
Quantities in parentheses
indicate one standard deviation uncertainty as calculated from five time
averages. To convert into atomic displacements, multiply by the displacement patterns described in
\tab{modes}.  Root-mean-square fluctuations for the other modes as a function of temperature
can be found in the Supporting Information Table S2.
}
\begin{tabular}{lccc}
\\
\hline
Mode $\mu$ &        & $\sigma({\mu})$ &  \\
           &  300 K & 600 K                 & 900 K \\
t$_{\rm 1u}^{(1)}(x;y)$ & 0.0138(1)  & 0.0207(4)  & 0.0294(3) \\
t$_{\rm 1u}^{(1)}(z)$   & 0.0592(1)  & 0.0495(4)  & 0.0294(3) \\
\hline
\label{tab:pbtio3b}
\end{tabular} \end{table}

The average crystallographic structures found in the AIMD simulations
are given in able~\ref{tab:pbtio3a}, 
while the root mean square fluctuations of the important
t$_{\rm 1u}^{(1)}$ modes are given in Table~\ref{tab:pbtio3b}
An example of how the local t$_{\rm 1u}^{(1)}(z)$ coordinate
for one specific Ti ion varies over time for different temperatures
is shown in \fig{tfluck}.  Normalization of the 
t$_{\rm 1u}^{(1)}(z)$ displacement pattern (\tab{modes}) means that the
displacement of the Ti relative to the neighboring oxygens in
the $\pm z$ directions is $3/\sqrt{6} \times c$
times the amplitude shown in this figure.

\begin{figure}[h]
\includegraphics[width=85mm]{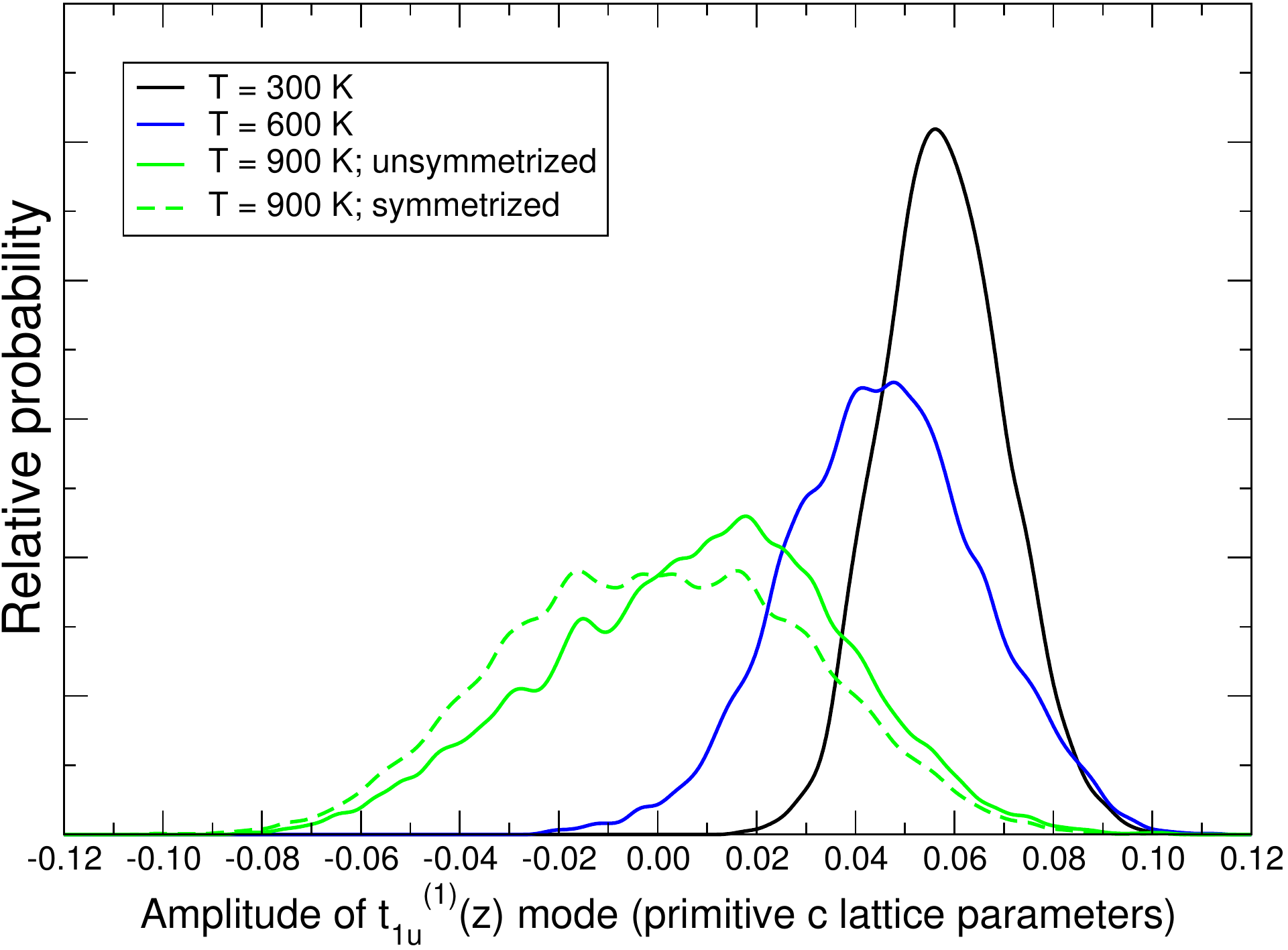}
\caption{Histograms of amplitude of t$_{\rm 1u}^{(1)}$(z) fluctuation
of Ti ions against their $\pm z$ O neighbors in PbTiO$_3$ versus temperature,
as calculated via ab initio molecular dynamics.  The raw data for $T$ = 900 K is
shown, as well as the data symmetrized about zero.}
\label{fig:histos}
\end{figure}

 A histogram of the amplitudes of the t$_{\rm 1u}^{(1)}(z)$ fluctuations of 
{\em all} Ti atoms versus temperature is shown in \fig{histos}.
For $T$ = 900 K, we additionally show the histogram of displacements 
obtained by symmetrizing the data over $\pm z$.  The asymmetry in
the ``raw" data is attributed to insufficient averaging time in the 2.5 ps
AIMD run.  Interestingly, at all temperatures investigated, the peak amplitude of 
the t$_{\rm 1u}^{(1)}$ fluctuation is similar: about 0.1, or about a {0.5~\AA}
relative displacement of Ti against O along the $z$ direction.

There are two possibilities for the behavior in the cubic phase above the  
ferroelectric transition at $T$ = 763 K:  if the amplitude peaks at
zero, that indicates a displacive nature to the Ti off-centering in
the ferroelectric transition; if the amplitude has two peaks at nonzero
values, that indicates an order-disorder nature of Ti off-centering.
Our present results for $T$ = 900 K can not clearly distinguish between the 
two cases.   It would be useful to repeat the AIMD calculation for temperatures 
closer to the transition temperature, but the correlation length and time of
the Ti off-centering fluctuation are expected to increase as the phase
transition is approached\cite{Hohenberg1977,Rabe1998}, requiring larger
simulations cells and and times for convergence.  Our AIMD simulations used
80-atom cells. Classical molecular dynamics based on a bond-valence-sum model 
for PbTiO$_3$ has recently been used to simulate cells with up to 2560 atoms\cite{Liu2013}. 
Determining the fluctuations of the t$_{\rm 1u}^{(1)}$ modes via
the classical MD approach could help clarify to what extent the root-mean 
square local fluctuations are affected by longer-range correlations near the transition temperature.

Experimentally, near-edge measurements 
reported by Yoshiasa~{\em{et al.}}~\cite{Yoshiasa2016} on a {\PTO} powder sample show
a nearly continuous evolution of the intensity of the E$_{\rm g}$-derived peak
near the transition temperature, suggesting a smooth trend for
the root-mean-square Ti off-centering vs temperature, but measurement 
of a root-mean-square off-centering by itself does not show whether the underlying
distribution has one peak or two.
Other experimental\cite{Sicron1994} 
and computational\cite{Fang2015} studies of {\PTO} support an order-disorder 
nature of the phase transition in {\PTO}.  Neither of these studies
measured precisely the same ``local order parameter" as the AIMD 
calculations in this work.   In light of our results, this raises the 
intriguing question of whether different possible local order parameters 
for the tetragonal-cubic transition in {\PTO} could give different answers
for the displacive vs order-disorder nature of the transition.

\begin{table}
\caption{Structures used in the Bethe-Salpeter equations to compute near-edge spectra in {\PTO}.
These are {\em not} the average crystallographic structures at the corresponding temperature,
but are specifically created such that Ti is off-centered along each coordinate by the root mean
square fluctuation as calculated by AIMD simulations.}
\begin{tabular}{lccc}
\\
\hline \hline
\\
& & $T$=300~K: \\
{\bf{Ion}} & {\bf{\em x}} & {\bf{\em y}} & {\bf{\em z}} \\ \hline
Pb & 0.0000 & 0.0000 & 0.1056 \\
Ti & 0.5113 & 0.5113 & 0.5483 \\
O  & 0.9944 & 0.5000 & 0.4982 \\
O  & 0.5000 & 0.9944 & 0.4982 \\
O  & 0.5000 & 0.5000 & 0.9758 \\ \hline
\\
& & $T$=600~K: \\
{\bf{Ion}} & {\bf{\em x}} & {\bf{\em y}} & {\bf{\em z}} \\ \hline
Pb & 0.0000 & 0.0000 & 0.0739 \\
Ti & 0.5169 & 0.5169 & 0.5404 \\
O  & 0.9915 & 0.5000 & 0.4993 \\
O  & 0.5000 & 0.9915 & 0.4993 \\
O  & 0.5000 & 0.5000 & 0.9798 \\ \hline
\\
& & $T$=900~K: \\
{\bf{Ion}} & {\bf{\em x}} & {\bf{\em y}} & {\bf{\em z}} \\ \hline
Pb & 0.0000 & 0.0000 & 0.0000 \\
Ti & 0.5240 & 0.5240 & 0.5240 \\
O  & 0.9880 & 0.5000 & 0.5000 \\
O  & 0.5000 & 0.9880 & 0.5000 \\
O  & 0.5000 & 0.5000 & 0.9880 \\ \hline
\\
\hline \hline
\label{tab:wyckparam}
\end{tabular}
\end{table}


\begin{figure}[h]
\includegraphics[width=85mm]{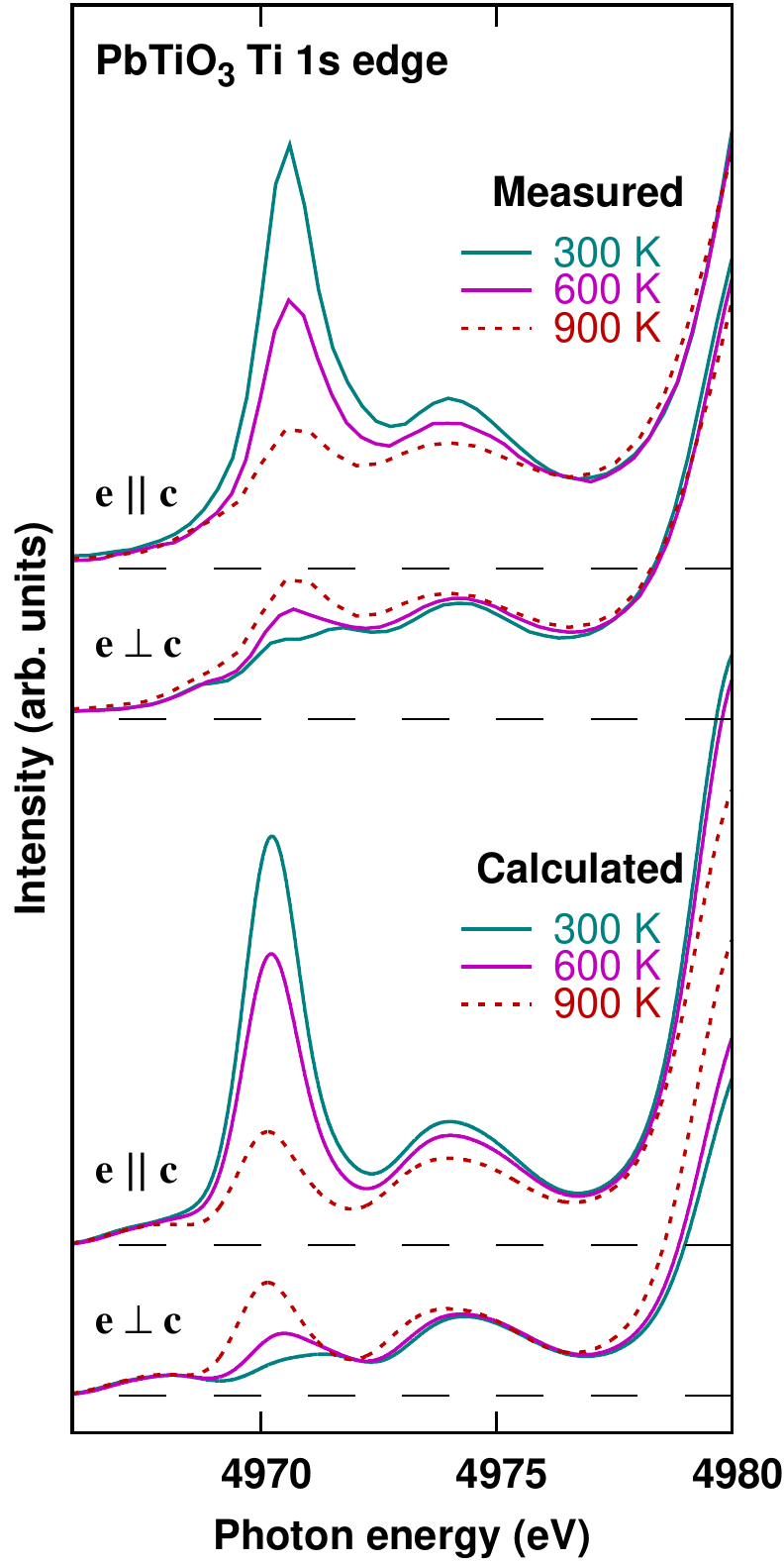}
\caption{Ti 1s near-edge spectrum emphasizing the pre-edge region for {\PTO} at three temperatures,
as measured and calculated.}
\label{fig:ptobse}
\end{figure}

Plugging the AIMD fluctuations for PbTiO$_3$ (\tab{pbtio3b} into the
general expression formula for the effective five-atom cell for BSE
calculations (\tab{nexafscell}), yields the temperature-dependent
effective structures given in~\tab{wyckparam}.
Based on these effective structures, computed near-edge spectra for each 
x-ray photon geometry are presented in \fig{ptobse} for three temperatures for {\PTO} and 
compared to measured spectra.  
The spectra are vertically offset for presentation, with the baseline 
indicated by a dashed line in each case.    
It should be noted that a rigid shift was required to energetically 
align the spectra, which is standard practice for calculated core excitation spectra.  
The peak that betrays the T$_{\rm 2g}$-derived core hole-exciton levels near 4968.5~eV 
shows minimal dependence on either temperature or x-ray photon geometry.  
On the other hand, the corresponding peak for E$_{\rm g}$-derived levels near 4970.5~eV 
shows a strong dependence 
on temperature.  
For 
${\bf e} \perp (001)$, 
it grows in strength with temperature because of 
admixture of the Ti 4p and Ti 3d$_{x^2-y^2}$ levels because of fluctuations of the 
Ti $x-$ and $y-$coordinates.  
However, this peak becomes particularly pronounced for ${\bf e} \parallel (001)$ 
at temperatures below the transition temperature, when Ti off-centering is largest, 
especially at lower temperatures.  
A broad feature around 4974~eV appears to be enhanced 
along with the E$_{\rm g}$-derived feature.  
Off-centering of the Ti affects hybridization of Ti 4p and O 2p states.  
In the unoccupied bands, there should be a node in such an antibonding state 
between the Ti and O atoms, and the weight of the Ti 4p state on the Ti site can 
be enhanced because of increasing localization 
arising from the node and foreshortened bond length~\cite{Vedrinskii1998,deGroot2009}.     

\begin{figure}[h]
\includegraphics[width=85mm]{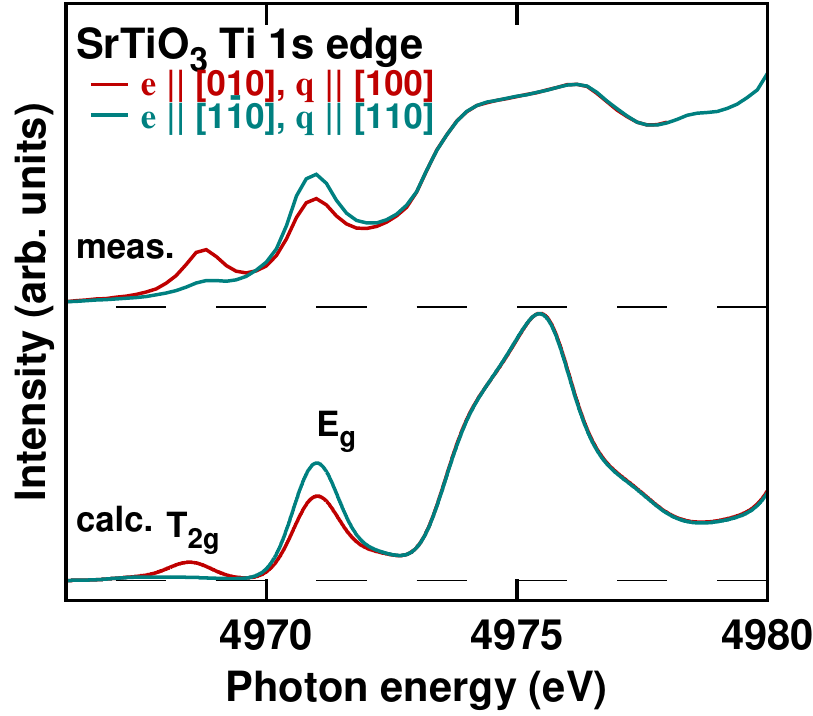}
\caption{Ti 1s near-edge spectrum emphasizing the pre-edge region for {\STO} at 293~K,
as measured and calculated.}
\label{fig:stobse}
\end{figure}

In addition to {\PTO}, we show results for {\STO} obtained at $T=$293~K~\cite{Woicik2007}.  
In this case, the two x-ray photon geometries that were considered were  
(1.) ${\bf{q}} \parallel (1,1,0)$ and $ {\bf{e}} \parallel (1,-1,0)$, and 
(2.) ${\bf{q}} \parallel (1,0,0)$ and $ {\bf{e}} \parallel (0,1,0)$.   
Na\"\i vely, the first geometry should make transitions to 
Ti 3d$_{x^2-y^2}$-derived core hole-exciton levels observable, 
though not transitions to Ti 3d$_{xy}$-derived core hole-exciton levels.  
For the second geometry, exactly the opposite should hold true.  In fact, 
while the selection rule works nearly perfectly for the $T_{\rm 2g}$-derived peak, 
the E$_{\rm g}$-derived peak is always visible with similar intensity in either case.  
We performed similar calculations in {\STO} as described for {\PTO}.  
AIMD simulations {\STO} at room temperature give fluctuations $\sigma(t_{1u}^{(1)}) 
= 0.0141$ along each Cartesian direction.
The resulting spectra are shown in \fig{stobse}.  
Good agreement between the theoretical and experimental results is
again observed, and the figure suggests similar breakdowns of 
selection rules being observed and calculated.  
Violations of selection rules have also been found that allow $s \rightarrow s$ transitions 
in x-ray absorption because of lattice vibrations, 
which has been reported by others~\cite{Manuel2012,Pascal2014,Vinson2014,Nemausat2017}.  
We also note that the exaggerated intensity of the features near 4975~eV in 
SrTiO$_3$ could result from the BSE calculation’s failure to account for charge-transfer processes 
that should appear like more effect screening of the Ti 1s hole when the excited 
core electrons leaves the Ti site~\cite{Woicik2015}.

\noindent
\section*{Conclusions}

This work considers Ti 1s near-edge spectroscopic signatures 
of atomic displacement from those of the ideal perovskite structure 
in {\PTO} (and {\STO}) for a variety of 
sample temperatures and experimental geometries.  
Using a combination of AIMD and BSE calculations, we have found that 
only one type of mode has a strong spectroscopic signature, 
namely the $t_{\rm 1u}^{(1)}$ degrees of freedom involving 
the Ti$^{4+}$ moving relative to its axial O neighbors.
This permits having the BSE calculations 
sample only a few frozen configurations of atomic coordinates 
to estimate effects on spectra accurately.
To a good approximation, 
the strong peak around 4970.5~eV varies as the mean square
displacement of this mode, 
and therefore does not vanish 
above the ferroelectric-paraelectric transition
temperature, even though 
the average structure has Ti at the center of its oxygen cage.

Our results show that AIMD and BSE 
can be combined to successfully demonstrate the local
fluctuations of Ti in {\PTO} and suggest that similar analysis 
could be fruitfully applied in other systems
such as relaxors.
We have not attempted to consider cage-rotation effects at 
low temperature in {\STO}.  In that case, the associated 
modes could be both low-frequency and considerably more anharmonic, 
which might require extensions beyond the present analysis.   
In earlier work, a single, consistent picture could be constructed 
regarding crystal structure and local atom geometries based on several methods, 
including x-ray diffraction, x-ray absorption, density-functional theory calculations of ground state structures, 
and the degree and directionality of ferroelectric polarization and/or local 
off-centering in several perovskites~\cite{Woicik2007,Woicik2017}. 
This work now incorporates molecular dynamics into the suite of collectively consistent 
and compatible methodologies in the same systems.


\begin{thebibliography}{51}
\bibitem [{dis()}]{disclaimer}%
  \BibitemOpen
  \href@noop {} {}\bibinfo {note} {Certain commercial software is identified in
  this paper to adequately describe the methodology used. Such identification
  does not imply recommendation or endorsement by the National Institute of
  Standards and Technology, nor does it imply that the software identified is
  necessarily the best available for the purpose.}\BibitemShut {Stop}%
\end{thebibliography}
\end{document}